\documentclass[12pt]{iopart}
\input epsf

\usepackage{amssymb}

\usepackage[dvips]{color}

\newcounter{append}
\def\moyenne#1{\langle #1 \rangle}

\def\LL{{\cal L}}

\def\ggg{{\bf g}}

\begin{document}
\jl{1}

\title{Network Evolution Induced by the Dynamical Rules of Two Populations}
\author{Thierry Platini and  R.K.P. Zia
\footnote{Corresponding author: {\tt platini@vt.edu}}}
\address{Department of Physics, Virginia Polytechnic Institute and State University, Blacksburg, Virginia 24061-0435, USA}
\date{\tt \today}
\begin{abstract}
We study the dynamical properties of a finite dynamical network composed of two interacting populations, namely; extrovert ($a$) and introvert ($b$). In our model, each group is characterized by its size ($N_a$ and $N_b$) and preferred degree ($\kappa_a$ and $\kappa_b\ll\kappa_a$). The network dynamics is governed by the competing microscopic rules of each population that consist of the creation and destruction of links. Starting from an unconnected network, we give a detailed analysis of the mean field approach which is compared to  Monte Carlo simulation data. The time evolution of the restricted degrees $\moyenne{k_{bb}}$ and $\moyenne{k_{ab}}$ presents three time regimes and a non monotonic behavior well captured by our theory. Surprisingly, when the population size are equal $N_a=N_b$, the ratio of the restricted degree $\theta_0=\moyenne{k_{ab}}/\moyenne{k_{bb}}$ appears to be an integer in the asymptotic limits of the three time regimes. For early times (defined by $t<t_1=\kappa_b$) the total number of links presents a linear evolution, where the two populations are indistinguishable and where $\theta_0=1$. Interestingly, in the intermediate time regime (defined for $t_1<t<t_2\propto\kappa_a$ and for which $\theta_0=5$), the system reaches a transient stationary state, where the number of contacts among introverts remains constant while the number of connections is increasing linearly in the extrovert population. Finally, due to the competing dynamics, the network presents a frustrated stationary state characterized by a ratio $\theta_0=3$.
\end{abstract}
\section{Introduction}
In the last decade, it became clear that the statistical properties of natural or artificial networks differ most of the time from the properties of random graphs. Nowadays, it is well known that the formation and the dynamics of complex networks are far from random but are governed by underlying organizing principles. Some of the most famous networks are: the living cells (formed by chemical and chemical reactions), the internet, the world wide web, power grids and social networks. See \cite{Strogatz-2001,Albert-Barabasi-2002,Dorogovtsev-2002,Newman-2003,Boccaletti-2006,Costa-2007} for review articles. In addition, it appears that networks are an interesting support for various dynamical processes where the topology should play a crucial role. As an example, the random walk \cite{Huberman-1998,Kleinberg-2000,Adamic-2001,Bilke-Peterson-2001,Burda-2001,Noh-2004} or the process of synchronization in complex networks \cite {Wu-2007,Arenas-2008} has been some of the favorite playgrounds of the scientific community. One should also mention the spread/diffusion of ideas, innovations or viruses \cite{Coleman-Menzel-Katz-1957,Valente-1995} that are of particular importance. These dynamical problems have been studied on regular \cite{Kauffman-1993,Keeling-1999}, random \cite{Solomonoff-Rapoport-1951,Rapoport-1957,Weigt-Hartmann-2001}, small world \cite{Moukarzel-1999,Newman-Watts-1999a,Newman-Watts-1999b,Moore-Newman-2000a,Moore-Newman-2000b, Newman-Moore-Watts-2000, Kuperman-Abramson-2001} and scale free networks \cite{Johansen-Sornette-2000,Tadic-2001}.\\ 
A challenging task, in the characterization of the network topology is the detection of sub-units \cite{Blatt-1996,Zhou-2003,Reichardt-2004,Newman-2004,Newman-2004b,Palla-2005,Derenyi-2005}, called clusters, modules or communities. These are the signature of the hierarchical nature of complex systems \cite{Ravasz-2002, Vicsek-2002,Guimera-2003}. In the case of social networks \cite{Watts-2002N} and in the framework of epidemiology, the spread of an infectious disease within a community and its transmission between communities is a key to understand and prevent pandemic outbreaks. Interestingly, while for random graphs, epidemiological models present a disease free state below a critical infection rate, it has been shown, for scale free network, that any spreading rate leads to the infection of the whole network \cite{Pastor-Vespignani-2001a,Pastor-Vespignani-2001b}. In an effort to build realistic models, epidemiologist focused on understanding the robustness of the results to the model assumptions \cite{Eubank-2007}. Finding out the simplest microscopic rules of a social network is a difficult task and is strongly dependent of the population under consideration. Networks that include social communities, social distance and group structures have been proposed \cite{Noh-2005,Boguna-2004}. \\
Motivated by the sociological interaction amongst extroverts and introverts, we analyze the dynamical properties of a finite size network composed of two populations. In an epidemiological context, we believe that the daily network of interactions is not necessarily scale free. We propose an alternative dynamics of the preferential attachment rule, where the populations are characterized by a `preferred degree', respectively $\kappa_{a}$ and $\kappa_{b}$ for extroverts and introverts. In our model, each individual tries to reach the `preferred degree' through the possibility to create or destroy bonds. The succession of these microscopic events governs the dynamics of the entire network. Evidently, the growth of the network is subject to the constraint imposed by the preferred degrees. Such constraints are known to lead to single scale networks \cite{Amaral-2000}. Examples are found in actors collaborative networks which present a exponential cutoff of the degree distribution. \\
We study the time dependence of such a network, starting with {\em no contacts} and evolving through birth, growth and relaxation. We find three regimes separated by two characteristic times $t_1$ and $t_2$ associated to the preferred degree $\kappa_{b}$ and $\kappa_{a}$, respectively.  In the early time regime, the two populations are indistinguishable while the second time regime is understood as a relaxation towards a transient stationary state. In this state, the number of links between extroverts is linearly increasing while the number of introvert-introvert and introvert-extrovert links are constant, depending on the difference of the population sizes. In the last time regime ($t_2<t$) the system relaxes towards the final, stationary state. Not surprisingly, the competing dynamics between the two populations induced frustration in the stationary state. This results simultaneously of the creation of cross-population links from the extrovert population and of their destruction by the introverts.\\
The paper is organized as follows: we start, in section 2, with a presentation of the model, introducing the probability $c$ to create a link as a function of the system parameters. In section 3, we give successively a full and coarse-grained description of the dynamics of our network through the master equation defined in the complete and reduced  configuration space. To pursue our analysis further, we derive, for finite system, the mean field equations describing the time evolution of the average degree of each population. In section 4, starting from an empty network we study the evolution of the network through the early, intermediate and late time regimes. For various parameter values, the prediction of the mean field theory (in the limit $N\rightarrow\infty$) are compare to Monte Carlo simulation data. The limitations of our approach are presented in section 5, especially in the intermediate time regime where the system presents unexpected finite size effects. We summarize our results in the last section and discuss possible extension of our study.

\section{The model}
Generally, a realistic network is composed of an arbitrary number of populations or groups. In our work, motivated by the competing aspect of different populations, we focus our attention on the interaction between just two groups: extrovert and introvert individuals. The model under consideration is composed of $N$ nodes divided in two populations, respectively $a$ (extrovert) and $b$ (introvert), of size $N_a$ and $N_b$ with $N=N_a+N_b$ being constant. Each group is characterized by its preferred degree $\kappa_a$ and $\kappa _b$. Extrovert and introvert individuals prefer to have respectively many and few contacts, which is modeled by imposing $\kappa _a\gg \kappa _b$. It is the succession of local events, such as creation or destruction of edges by each individual, that governs the dynamics of the entire network. Starting with a totally disconnected network (i.e., zero contacts), we study the growth of connections induced by the microscopic rules. Of course, more sophisticated versions, including the birth and death process of nodes or ``mutation'' from one group to the other, can be considered.\\

In our simulations, one time-step is defined by the successive random selection of $N$ individuals where each node is updated once, according to the rules detailed below. Defining 
\begin{eqnarray}
\Delta \equiv (N_a-N_b)/N, 
\end{eqnarray}
the probability to select a node of the population $a$ is given by $(1+\Delta )/2$ (and $(1-\Delta )/2$ for $b$) such that on average, all nodes are updated once in each time-step. We will label a node by $j$, denote its social group by ${\bf g}(j)$ (${\bf g}(j)=a,b$), and its actual degree by $k_j$. When a node is chosen, we will create or destroy a link attached to it, with respectively probability $c$ or $1-c$. For simplicity, we will choose the same structure of the probability of creation $c$ for the two groups. The only difference being the preferred degree, $\kappa _{{\bf g}(j)}$, which enters through the dependence on $k_j$ in terms of the ratio 
\begin{eqnarray}
r_j\equiv k_j/\kappa _{{\bf g}(j)}\,\,. 
\end{eqnarray}
Clearly, we must insist that $c=1$ if the degree $k_j=0$, since there can be no more links to destroy. On the other hand, a natural condition is $c\rightarrow 0$ if $k_j\gg \kappa _{{\bf g}(j)}$. The crossover from $1$ to $0$ can be characterized by some width, which we also introduce as a parameter: $\sigma$. We may regard this as a measure of the `flexibility' or `tolerance' level of the population, with larger $\sigma $ corresponding to {\em higher} `rigidity' or {\em lower} ability to tolerate a degree $k$ far from the preferred value.
Specifically, we choose 
\begin{eqnarray}
c(r)=\frac 1{1+r^\sigma },  \label{ProbaC}
\end{eqnarray}
and denote, for the node $j$, 
\begin{eqnarray}
c_j\equiv c(r_j). 
\end{eqnarray}
as the probability to create a new link with a randomly chosen individual who is not already in contact with $j$. On the other hand, with probability $d_j=1-c_j$, we randomly choose one of its existing links to destroy. After the update, its degree $k_j^{\prime }$ is simply $k_j\pm 1$, a restriction which can be easily lifted in future studies.\\

Clearly, form equation (\ref{ProbaC}), the `width' of the crossover region decreases with $1/\sigma $, so that $c(r)$ approaches the Heaviside function $H(1-r)$ in the limit $\sigma\rightarrow \infty $. Much of our analysis will be based on this limit and its associated predictions for the system's behavior are expected to be universal. Note that the probability $c$ is the familiar Fermi-Dirac distribution, by identifying $k,\kappa ,\sigma $ with $e^\epsilon ,e^\mu ,\beta $. In our context, it has another advantage, namely, the ratio of destruction to creation probabilities is a simple power of $r$: $d/c=r^\sigma $. Other choices for $c$ are of course possible. However, we believe the detailed specifics of $c$ will lead to only non-essential properties displayed by the system. For most of our simulations we choose: $N=1000$, $\Delta =0$, $\kappa_a=100$, $\kappa_b=10$, and
various $\sigma $'s in the interval $\left[2,100\right]$.
\section{Master equations and a mean-field approach}
\subsection{Full description of the system}
Such interwoven systems are particularly well described by the powerful approach and tools of statistical mechanics. It is clear that only few exact results can be extracted and that an analytical description remains challenging. The size of the configuration space illustrates perfectly the complexity of networks related problems. We start by defining the contact variables $\mu _{i,j}$ between the nodes $i$ and $j$, such that $\mu_{i,j}=1 $ if a link connects these points and $\mu _{i,j}=0$ otherwise. As we exclude `self-links' in our model, $\mu _{i,i}\equiv 0$ at all times. Thus, a configuration, $X$, is uniquely labeled by the symmetric (off diagonal) `matrix' $\mu _{i,j}$. Since there are ${\cal L}=N(N-1)/2$ elements in such matrices, there are $2^{\LL}$ possible configurations (i.e., $X=1,2,...2^{\LL}$). The master equation of the probability $P_t(X)$ to find the system in a particular state at a time $t$, has the usual
structure 
\begin{eqnarray}
P_{t+1}(X)=\sum_{Y\ne X}W_X^YP_t(Y),
\end{eqnarray}
where $W_X^Y$ is the probability to make a transition from $Y$ to $X$. Note that here, $t+1$ is not a time-step as defined above, but a single attempt to update the network. Since one link is always created or destroyed, the probability for any $X$ to remain unchanged is zero. Of course, it is possible to write an explicit expression of the probabilities $W_X^Y$ in terms of the elements of the set $\left\{ c_j\right\}$. However, this approach is intractable and so, we will not pursue it further. We will consider simpler approximations below.
\subsection{Reduced phase space}
\begin{figure}[h]
\begin{center}
\epsfxsize=10cm
\mbox{\epsfbox{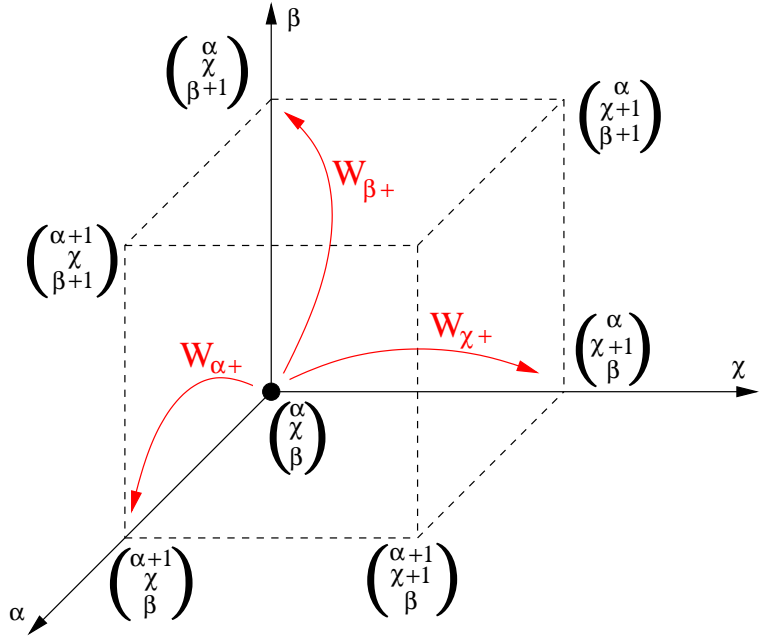}}
\caption{Reduced phase space of the network}
\label{CUBE}
\end{center}
\end{figure}

Instead of the detailed picture above, we believe most of the essentials of this network will be captured by a coarse-grained view, namely, a $3$ dimensional phase space spanned by $\alpha $ ($\beta $), the total number of links between extroverts (introverts), and $\chi $, the total number of links {\em between the two groups} (see Fig. $\ref{CUBE}$). Defining $P_t(\alpha ,\chi ,\beta )$ as the probability to find the system in a macrostate defined by the triplet $(\alpha ,\chi ,\beta )$, its master equation takes the following form 
\begin{eqnarray}
P_{t+1}(\alpha ,\chi ,\beta ) &=&\sum_{i=\pm 1}P_t(\alpha +i,\chi ,\beta
)W_{\alpha ,\mp }(\alpha +i,\chi )  \label{MasterEq} \\
&\hphantom{A}&  \nonumber \\
&+&\sum_{i=\pm 1}P_t(\alpha ,\chi +i,\beta )W_{\chi ,\mp }(\alpha ,\chi
+i,\beta )  \nonumber \\
&\hphantom{A}&  \nonumber \\
&+&\sum_{i=\pm 1}P_t(\alpha ,\chi ,\beta +i)W_{\beta ,\mp }(\chi ,\beta +i), 
\nonumber
\end{eqnarray}
where $W_{\nu ,+}$ and $W_{\nu ,-}$ are the probabilities to create and destroy a link of type $\nu $ ($\nu =\alpha ,\chi ,\beta $). Even in this simplified picture, the exact transition probabilities ($W_{\nu,\pm}$) during one time step are not easy to write, since that would require the knowledge of how the $\alpha ,\chi$ and $\beta $ links are distributed in the system. To proceed, we assume that all the nodes of the same population have the same degree, so that only their {\em averages}, denoted by $k_{{\bf g}}$ (${\bf g}=a,b$), enter into our $W$. Clearly, the degrees $k_a$ and $k_b$ receive contributions from links within the groups $\alpha$ and $\beta$ and those across groups ($\chi $). Specifically, defining the restricted degree $k_{{\bf g}{\bf g}^{\prime }}$ by 
\begin{eqnarray}
k_{aa}\equiv \frac{2\alpha }{N_a},\ \ k_{ab}\equiv \frac{2\chi}{N},\ \
k_{bb}\equiv \frac{2\beta }{N_b},
\end{eqnarray}
we have 
\begin{equation}
k_a=k_{aa}+\frac{k_{ab}}{1+\Delta },\quad\textrm{and} \quad k_b=k_{bb}+\frac{k_{ab}}{1-\Delta }.
\label{kakb}
\end{equation}
From these, we write the ratios $r_{{\bf g}}=k_{{\bf g}}/\kappa _{{\bf g}}$ and construct the creation probabilities: 
\begin{eqnarray}
c_{{\bf g}}=\frac 1{1+r_{{\bf g}}^\sigma }.
\end{eqnarray}
Let us emphasize, instead of having $N$ individual $c_j$'s, we deal with only two ``group'' $c$'s here: $c_a$ and $c_b$. Next, noting that the creation of a link by an individual requires the random selection of another node {\em not already connected}, we will approximate that probability by 
\begin{eqnarray}
p_{{\bf g}}=\frac{N_{{\bf g}}-1-k_{{\bf g}{\bf g}}}N,\ \ {\bf g}=a,b.
\end{eqnarray}
Thus, within the group $a$ and $b$, the probabilities $W_{\alpha,+}$ and $W_{\beta,+}$ are given by 
\begin{eqnarray}
W_{\alpha ,+}(\alpha ,\chi )=\frac{N_a}Np_ac_a,\quad\textrm{and}\quad W_{\beta ,+}(\chi ,\beta )=\frac{N_b}Np_bc_b.
\end{eqnarray}
Note that the probabilities ($W$) depend on only two of the three variables ($\alpha $,$\chi$,$\beta $). Following the same lines and defining $d_{{\bf g}}\equiv1-c_{{\bf g}}$, we arrive at 
\begin{eqnarray}
W_{\alpha ,-}(\alpha ,\chi )=\frac{N_a}N\frac{k_{aa}}{k_a}d_a,\quad\textrm{and}\quad W_{\beta
,-}(\chi ,\beta )=\frac{N_b}N\frac{k_{bb}}{k_b}d_b,
\end{eqnarray}
where $k_{{\bf g}{\bf g}}/k_{{\bf g}}$ is the probability to pick, among the existing links, a link connecting a node of the same population. Finally, for the creation and destruction of cross-links, we have 
\begin{eqnarray}
W_{\chi ,+}(\alpha ,\chi ,\beta ) &=&\frac{N_a}N\frac{N_b-k_{ab}}Nc_a+\frac{N_b}N\frac{N_a-k_{ba}}Nc_b, \\
W_{\chi ,-}(\alpha ,\chi ,\beta ) &=&\frac{N_a}N\frac{k_{ab}}{k_a}d_a+\frac{N_b}N\frac{k_{ba}}{k_b}d_b.
\end{eqnarray}
As usual, if $P_t(\alpha ,\chi ,\beta )$ is known, it can be used to find the averages of all observable quantities. A relevant example is given by the average degree associated with each population at any time $t$: 
\begin{equation}
\left\langle k_{{\bf g}}\right\rangle _t=\sum_{\alpha,\chi,\beta} k_{{\bf g}}P_t(\alpha ,\chi,\beta ).
\end{equation}
For convenience, we will drop the subscript in $\langle \bullet \rangle _t$ below, except when its presence is needed for clarity. We will also discuss time independent (or almost so) quantities, which will be denoted by a suitable superscript (e.g., $\langle \bullet \rangle ^{*}$).
\subsection{A mean-field approach}
Even in the reduced phase space, the master equation ($\ref{MasterEq}$) cannot be solved exactly. To gain some insight into our system, we proceed with a mean-field approximation (i.e., replacing averages of functions by functions of averages). In this process, we will often encounter creation probabilities that depend on $\left\langle k_{{\bf g}}\right\rangle ^\sigma $. Therefore, let us define for convenience
\begin{eqnarray}
f\equiv \frac 1{1+\left\langle k_a\right\rangle ^\sigma /\kappa _a^\sigma },\quad\textrm{and}\quad\quad h\equiv \frac 1{1+\left\langle k_b\right\rangle ^\sigma /\kappa_b^\sigma }.
  \label{fh}
\end{eqnarray}
From the exact equation $\partial _t\langle \nu\rangle =\langle W_{\nu ,+}\rangle -\langle W_{\nu ,-}\rangle $, we find 
\begin{eqnarray}
\partial _t\langle \alpha \rangle =\frac{N_a}N\frac{N_a-1-\langle k_{aa}\rangle }Nf-\frac{N_a}N\frac{\langle k_{aa}\rangle }{\langle k_a\rangle }(1-f),
\end{eqnarray}
which, recalling $k_{aa}=2\alpha /N_a$ and rescaling $t\rightarrow Nt$, leads to
\begin{eqnarray}
\partial _t\langle k_{aa}\rangle =\left( 1+\Delta -2\frac{1+\langle k_{aa}\rangle }N\right) f-2\frac{\langle k_{aa}\rangle }{\langle k_a\rangle } (1-f),
\label{EQA1}
\end{eqnarray}
and similarly
\begin{eqnarray}
\partial _t\langle k_{bb}\rangle =\left( 1-\Delta -2\frac{1+\langle k_{bb}\rangle }N\right) h-2\frac{\langle k_{bb}\rangle }{\langle k_b\rangle } (1-h).
\label{EQA1b}
\end{eqnarray}
The equation for the restricted degree $\langle k_{ab}\rangle $ is also straightforward, though slightly different: 
\begin{equation}
\partial _t\langle k_{ab}\rangle =\left( \frac{1-\Delta ^2}2-\frac{\langle k_{ab}\rangle }N\right) (f+h)-\frac{\langle k_{ab}\rangle }{\langle k_a\rangle }(1-f)-\frac{\langle k_{ab}\rangle }{\langle k_b\rangle }(1-h).
\label{EQA2}
\end{equation}
With the equations
\begin{eqnarray}
\langle k_a\rangle =\langle k_{aa}\rangle +\langle k_{ab}\rangle /(1+\Delta),\ \ \langle k_b\rangle =\langle k_{bb}\rangle +\langle k_{ab}\rangle/(1-\Delta),
\end{eqnarray}
we have a closed set of three ordinary, nonlinear differential equations for three functions: $\langle k_{aa}\rangle ,\langle k_{ab}\rangle ,\langle k_{bb}\rangle $. Of course, we can make linear combinations of the above to write equations for $\left\langle k_a\right\rangle $ and $\left\langle k_b\right\rangle $. However, this two equations do not close since both involve $\langle k_{ab}\rangle $. One instructive combination is given by $\rho$
\begin{eqnarray}
\rho &\equiv &\frac{1+\Delta }{4}\langle k_{aa}\rangle +\frac{1}{2}\langle k_{ab}\rangle +\frac{1-\Delta }{4}\langle k_{bb}\rangle  \label{Q-def} \\ 
&=&\frac{1+\Delta }4\langle k_a\rangle +\frac{1-\Delta }4\langle k_b\rangle 
\nonumber
\end{eqnarray}
which is the related to the total number of links in the entire system by the relation $N\rho=\moyenne{\alpha}+\moyenne{\chi}+\moyenne{\beta}$. Its evolution is governed by 
\begin{eqnarray}
\partial _t\rho &=&\left[ \left( 1+\Delta \right) f+\left( 1-\Delta \right) h-1\right]  \label{dQ/dt}\nonumber \\
&&-\frac{1}{2N}\left[ \left( 1+\Delta \right) \left( 1+\langle k_a\rangle \right) f+\left( 1-\Delta \right) \left( 1+\langle k_b\rangle \right) h\right]  
\label{Q-1/N}.
\end{eqnarray}
A convenient simplification occurs if we take the limit $N\rightarrow \infty$, namely, 
\begin{eqnarray}
\partial _t\langle k_{aa}\rangle &=&\left( 1+\Delta \right) f-2\frac{\langle k_{aa}\rangle }{\langle k_a\rangle }(1-f),  \label{MEANFIELDX}\\
\partial _t\langle k_{bb}\rangle &=&\left( 1-\Delta \right) h-2\frac{\langle k_{bb}\rangle }{\langle k_b\rangle }(1-h), \label{MEANFIELDX2}\\
\partial _t\langle k_{ab}\rangle &=&\frac{1-\Delta ^2}2(f+h)-\langle k_{ab}\rangle \left[ \frac{1-f}{\langle k_a\rangle }+\frac{1-h}{\langle k_b\rangle }\right].
\label{MEANFIELDX3}
\end{eqnarray}
In the following, we will investigate the time evolution of these quantities, using both Monte Carlo simulations and the equations above.
\section{Time evolution}
\begin{figure}[h] 
\begin{center}
\epsfxsize=10cm
\mbox{\epsfbox{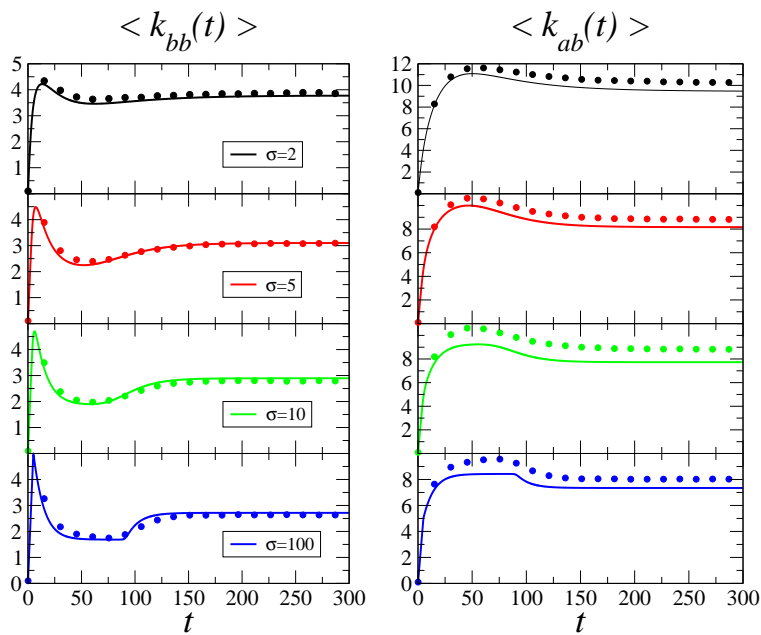}}
\caption{Comparison of the mean field predictions (lines) with the simulation results (symbols). The time evolution of the degrees $\langle k_{bb}\rangle (t)$ and  $\langle k_{ab}\rangle (t)$ are plotted for different values of $\sigma $, with $\kappa_a=100$, $\kappa _b=10$ and $\Delta =0$.}
\label{Graph1}
\end{center}
\end{figure}

In this section, we present the studies of how an initially empty network evolves to the steady state. Such a scenario may be realized when the population is formed from a collection of `complete strangers,' such as at the start of a cruise or the beginning of a school year (for, e.g., kindergarteners or college Freshmen). Simulating the
network according to the rules above, we discover three time regimes, as well as non monotonic behavior in the time evolution of the restricted degree $\langle k_{ab}\rangle _t$ and $\langle k_{bb}\rangle _t$. See figure ($\ref{Graph1}$). As we will show, the different regimes can be understood in terms of the mean-field approach. In particular, provided the $\kappa $'s are well separated, the characteristic times ($t_1$ and $t_2$) separating these regimes become quite well-defined for systems with large $\sigma $. Such cases correspond to a highly `rigid' population, for which both creation and destruction of links are highly probable even for small values of $\left| \langle k_{{\bf g}}\rangle _t-\kappa _{{\bf g}}\right| $. We will analyze the three regimes successively, for a range of values of the system parameters. 
\subsection{Early time regime}
\begin{figure}[h!] 
\begin{center}
\epsfxsize=10cm
\mbox{\epsfbox{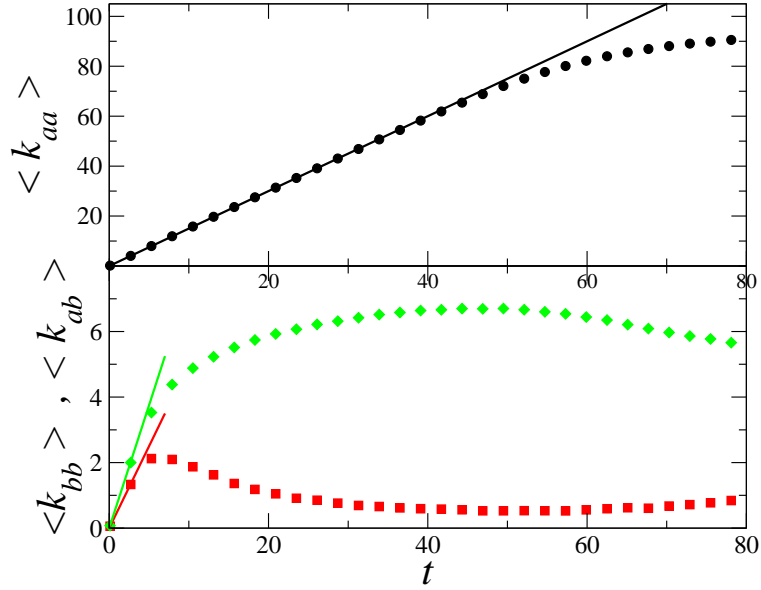}}
\caption{Time evolution of the average restricted degrees $\langle k_{aa}\rangle $ (circles), $\langle k_{bb}\rangle $ (squares) and $\langle k_{ab}\rangle $ (diamonds) in the early time regime. The data are extract from simulation (symbols) realized on a population of size $N=1000 $, after $10$ realizations, for $\kappa _a=100$, $\kappa _b=10$, $\Delta =1/2 $ and $\sigma =10$. The lines are given by the mean field results, equations ($\ref{Zxpetit}$).}
\label{SimulationShort}
\end{center}
\end{figure}

\begin{figure}[h!] 
\begin{center}
\epsfxsize=10cm
\mbox{\epsfbox{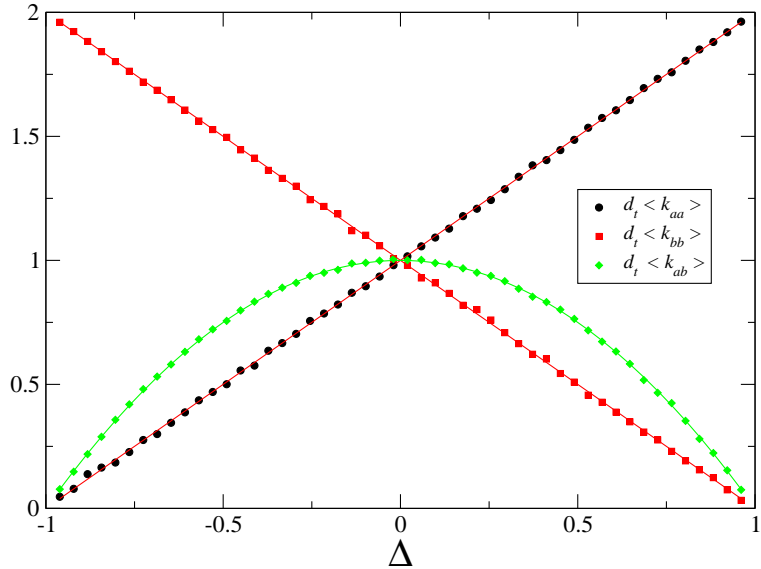}}
\caption{Slope of the average degrees $\langle k_{aa}\rangle $, $\langle k_{bb}\rangle $ and $\langle k_x\rangle $ at time $t=0$, as a function of the parameter $\Delta $. The data are extract from simulation (symbols) realized on a population of size $N=1000$, after $100$ realizations, for $\kappa _a=100$, $\kappa _b=10$ and $\sigma =10$. The lines are given by the mean field results, equations ($\ref{Zxpetit}$).}
\label{Slope}
\end{center}
\end{figure}

In the early time regime, the network is nearly empty, so that the probabilities of creation of links are close to unity. Then, the average number of links $\moyenne{\nu}$ ($\nu=\alpha,\chi,\beta$) grows linearly with $t$ and it is easy to find the slopes for each of the three classes of links. In the limit $\sigma \rightarrow \infty $, the probabilities of creation $c_{{\bf g}}$ become Heaviside functions and we can define the early time regime by the times $t<t_1$,
in which $c_{{\bf g}}=1$. Equation (\ref{Q-1/N}) simply leads to $\rho(t)=t$ while (\ref{MEANFIELDX}), (\ref{MEANFIELDX2}) and (\ref{MEANFIELDX3}) predict trivially 
\begin{equation}
\langle k_{aa}\rangle _t=(1+\Delta )t,\quad \langle k_{bb}\rangle
_t=(1-\Delta )t,\quad \langle k_{ab}\rangle _t=(1-\Delta ^2)t.
\label{Zxpetit}
\end{equation}
We can include finite population size effects and use ($\ref{EQA1}$-$\ref{EQA2}$) to obtain rounding of this linear growth. In this regime, we should note that $\theta_\Delta$ defined by the ratio $\moyenne{k_{ab}}/\moyenne{k_{bb}}$, is time-independent and given by
\begin{eqnarray}
\theta_\Delta={1+\Delta}, \, \forall t<t_1,
\end{eqnarray}
with $\theta_0$=1. Remarkably, even if the size of the populations are different ($\Delta \ne 0$), the growth of the degrees of the two groups are indistinguishable: 
\begin{eqnarray}
\langle k_a\rangle _t=\langle k_b\rangle _t=2t,
\end{eqnarray}
a result that breaks down for $t\simeq t_1$ and $\sigma =O\left( 1\right) $. Now, we see the meaning of $t_1$: It marks the time when the introverts reach their `tolerance' level, i.e., $\langle k_b\rangle _{t_1}=\kappa _b$. Thus, we have 
\begin{eqnarray}
t_1=\kappa _b/2.
\end{eqnarray}
If we include finite $N$ effects, then this becomes $-\frac N2\ln \left( 1-\frac{\kappa _b}{N-1}\right) $. In figure ($\ref{SimulationShort}$) and ($\ref{Slope}$) these results are compared to the Monte Carlo simulation data. The short time evolution of the restricted degrees, $\langle k_{{\bf gg}^{\prime }}\rangle _t$, are plotted for $\sigma =10$. The linear increase of the degrees is in perfect agreement with our predictions. The comparison for various values of $\Delta $ are shown in Fig ($\ref{Slope}$). In the latter figure, we plot the time derivative of the degrees at time $t=0$ ($\partial_t\langle k_{aa}\rangle |_{t=0}$, $\partial _t\langle k_{bb}\rangle |_{t=0}$ and $\partial _t\langle k_{ab}\rangle |_{t=0}$) as a function of $\Delta $. Agreement between theory and simulations is again excellent. 
\subsection{Intermediate time regime}
Similar to considerations above, let us define $t_2$ by the condition $\langle k_a\rangle _{t_2}=\kappa _a$ which satisfies the extrovert population. In the regime, $t_1\lesssim t\lesssim t_2$, with $\kappa _a>\kappa _b$ (and relatively small $\Delta $), the behavior of the system is characterized by two{\em \ competing} tendencies. The extroverts have yet to reach their preferred connectivity and so, continue to attempt creating links. Since some of these will be cross-links ($\chi $), the introverts find themselves having more connections than the preferred level and tend to destroy links. In the case $\kappa _b\ll \kappa _a$ , we can expect $t_1\ll t_2$ and the nature of this intermediate `state' is more exposed. In particular, a very large system (e.g., $N\gg \kappa _a$) may reach a `steady' state, in which the extroverts continue to create links at a constant rate. On average, a constant fraction of these new links will be cross-links, which are then destroyed by the introverts. In such a state, the rate of creation of cross-links by the extroverts balances the rate of destruction by the introverts. Thus, both restricted degree $\langle k_{ab}\rangle $ and $\langle k_{bb}\rangle $ remain essentially constant, values which will be denoted by $\langle k_{ab}\rangle ^{\star }$ and $\langle k_{bb}\rangle ^{\star }$. The continued growth in $\langle k_a\rangle $ must be entirely through that in $\langle k_{aa}\rangle$ (which is $\approx (1+\Delta )t$). Finally, the end of this regime is marked by the extroverts reaching their preferred $\kappa _a$. We can estimate $t_2$ through $\kappa _a=\langle k_a\rangle _{t_2}$ with
\begin{eqnarray}
\kappa _a=\langle k_{aa}\rangle _{t_2}+\frac{\langle k_{ab}\rangle ^{\star }}{1+\Delta }. 
\end{eqnarray}
We will show that $\langle k_{ab}\rangle ^{\star }=O\left( \kappa _b\right)$, and so, for $\kappa _b\ll \kappa _a$ , we find 
\begin{eqnarray}
t_2\cong \frac{\kappa _a}{1+\Delta }\,\,.
\end{eqnarray}
In the remainder of this subsection, we will demonstrate that this scenario is borne out, at least qualitatively.\\

Remaining with the simple (but extreme) case of $\sigma ,N\rightarrow \infty$ and approximating the creation probability for the extroverts by unity (i.e., $f=1$) leads to a consistent solution. By ignoring terms with $\left(1-f\right) $, $\langle k_{aa}\rangle$ formally decouples and we need to solve a system of only two variables, i.e., $\langle k_{bb}\rangle $ and $\langle k_{ab}\rangle $. Specifically, equations (\ref{MEANFIELDX}), (\ref{MEANFIELDX2}) and (\ref{MEANFIELDX3}) reduce to 
\begin{equation}
\partial _t\langle k_{aa}\rangle =\left( 1+\Delta \right)  \label{kaa eqn},
\end{equation}
and 
\begin{eqnarray}
\partial _t\langle k_{bb}\rangle &=&\left( 1-\Delta \right) h-2\frac{\langle
k_{bb}\rangle }{\langle k_b\rangle }(1-h),  \label{kbb eqn} \\
\partial _t\langle k_{ab}\rangle &=&\frac{1-\Delta ^2}2(1+h)+\frac{%
\left\langle k_{ab}\right\rangle }{\langle k_b\rangle }(1-h).
\label{kab eqn}
\end{eqnarray}
Since the dynamics of the extroverts decouples (except through an external, ``source-like'' term in the second equation), it is sensible to consider an equation for $\langle k_b\rangle $ in lieu of (\ref{kab eqn}): 
\begin{eqnarray}
\partial _t\langle k_b\rangle &=&\frac{1+\Delta }2+\frac{3-\Delta }2h-\left( 
\frac{\langle k_{bb}\rangle }{\langle k_b\rangle }+1\right) (1-h) \\
&=&\frac{-1+\Delta }2+\frac{5-\Delta }2h-\frac{\langle k_{bb}\rangle }{%
\langle k_b\rangle }(1-h).
\label{kb eqn}
\end{eqnarray}
Since $h$ is a function of $\langle k_b\rangle $ alone, the latter equation together with equation (\ref{kbb eqn}) form an explicitly closed set. At large times, we have the solution 
\begin{eqnarray}
\langle k_{aa}\rangle =\left( 1+\Delta \right) t,\quad \langle k_{bb}\rangle
\rightarrow \langle k_{bb}\rangle ^{\star },\quad \langle k_b\rangle
\rightarrow \langle k_b\rangle ^{\star },\quad h\rightarrow h^{\star }, 
\end{eqnarray}
where the superscript ($^{\star }$) denotes (quasi-) stationary values of the relevant quantities. In particular, eliminating the last terms from equations (\ref{kbb eqn}) and (\ref{kb eqn}), we easily find 
\begin{eqnarray}
h^{\star }=\frac{1-\Delta }4, 
\end{eqnarray}
and so, for $\sigma \gg 1$, 
\begin{eqnarray}
\langle k_b\rangle ^{\star }=\kappa _b. 
\end{eqnarray}
Notice that these results are `universal' in that they do not depend on the explicit form of the probability of creation (\ref{ProbaC}), but only on its being well approximated by a Heaviside function. The interpretation of the latter is particularly appealing. The introverts are satisfied, as far as their overall degrees are concerned. The only dynamics left in this regime is a redistribution, from cross-links ($\langle k_{ab}\rangle $) to intra-group links ($\langle k_{bb}\rangle $). Before the (quasi-) stationary state is reached, the increase of cross-links are, of course, driven by the extroverts (trackable to the first term in equation (\ref{kab eqn})).\\
The other (quasi-) stationary values follow readily: 
\begin{equation}
\langle k_{bb}\rangle ^{\star }=\frac{\kappa _b}2\frac{(1-\Delta )^2}{3+\Delta },\quad \langle k_{ab}\rangle ^{\star }=(5-\Delta )\langle k_{bb}\rangle ^{\star},
\label{kbb-star}
\end{equation}
with the ration $\theta_\Delta=\moyenne{k_{ab}}^{\star}/\moyenne{k_{bb}}^{\star}=5-\Delta$ and $\theta_0=5$. In the lowest panels of figure ($\ref{Graph1}$), for which $\Delta =0$, $\sigma =100$ and $N=1000$, this (quasi-) stationary state is well represented, in the range $50\lesssim t\lesssim 80$. The solid lines are results of numerical integration of the full set of equations (\ref{MEANFIELDX}), (\ref{MEANFIELDX2}) and (\ref{MEANFIELDX3}). They confirm this scenario quite well. An example is the ratio $\theta_0=\langle k_{ab}\rangle ^{\star }/\langle k_{bb}\rangle ^{\star }=5$. The dots are results from simulation. Clearly, there is also qualitatively good agreement, especially considering what a drastic approximation went into formulating this simple theory.
\subsection{Late time regime}
The onset of the last time regime is characterized by the following. While finally the extroverts reach (on average) their preferred degree $\kappa _a$, the system is not completely ``satisfied'' as the introverts have been driven, up to this time, by the zealous creation of cross-links from the extroverts. Thus, readjustments will take place, through mostly the destruction of cross-links. The decrease of the number of cross-links $\chi $ in turn affects the introverts, who compensates by creating more links. Again, the last panels of figure ($\ref{Graph1}$) showcase this scenario quite well, starting at $t\thicksim 80$. With both degrees being mostly satisfied, the two populations spend this final time period readjusting the distributions of cross-links and intra-group links, until the true stationary state is reached.\\

Let us denote the relevant quantities of the final state by a superscript ($^{*}$), and seek the solution to $\partial _t\langle k_{{\bf gg}^{\prime}}\rangle ^{*}=0$. Again, for simplicity, we present the results in the limiting case $\sigma ,N\rightarrow \infty $. First, considering (\ref{dQ/dt}), we find 
\begin{eqnarray}
\left( 1+\Delta \right) f^{*}+\left( 1-\Delta \right) h^{*}=1.
\label{EqualityCaCb}
\end{eqnarray}
This equation expresses a universal, and intuitively understandable, aspect of the stationary state. Dividing both sides by $2$, the left can be recognized as the probability to create a link in the system, regardless of which group the individual is a member. Therefore, equality (\ref{EqualityCaCb}) simply states that this probability must be $1/2$. The other equations for $f^{*}$ and $h^{*}$ can be obtained by rewriting the right hand sides of $\partial _t\langle k_{{\bf gg}}\rangle ^{*}=0$: 
\begin{eqnarray}
\left( 1+\Delta \right) f^{*}-2\left[ 1-\frac{\langle k_{ab}\rangle ^{*}%
}{\left( 1+\Delta \right) \langle k_a\rangle ^{*}}\right] (1-f^{*}) &=&0, \\
\left( 1-\Delta \right) h^{*}-2\left[ 1-\frac{\langle k_{ab}\rangle ^{*}%
}{\left( 1-\Delta \right) \langle k_b\rangle ^{*}}\right] (1-h^{*}) &=&0.
\end{eqnarray}
Since $f^{*}$ and $\langle k_a\rangle ^{*}$ are uniquely related to each other (and similarly for $h^{*}$ and $\langle k_b\rangle ^{*}$) one has
\begin{eqnarray}
\left( 1+\Delta \right) \langle k_a\rangle ^{*}\left[ \frac{\left( 1+\Delta
\right) f^{*}}{2(1-f^{*})}+1\right] =\langle k_{ab}\rangle ^{*}=\left(
1-\Delta \right) \langle k_b\rangle ^{*}\left[ \frac{\left( 1-\Delta \right)
h^{*}}{2(1-h^{*})}+1\right],\nonumber\\
\end{eqnarray}
which provides the necessary second equation between $f^{*}$ and $h^{*}$. Note that, like equation (\ref{EqualityCaCb}) this does not depend on the explicit functional forms of the probability $c_{{\bf g}}$. To continue further, we can recall the explicit forms, such as $\langle k_a\rangle ^{*}=\kappa _a\left(1/f^{*}-1\right) ^{1/\sigma }$, or  consider the limit $\sigma \rightarrow\infty$, in which we simply have, regardless of the details of $c_{{\bf g}}$; 
\begin{equation}
\langle k_{{\bf g}}\rangle ^{*}=\kappa _{{\bf g}}.
\end{equation}
Then, the only parameters which control the fixed point are 
\begin{equation}
R\equiv \kappa _b/\kappa _a\quad \textrm{and}\quad \Delta \equiv (N_a-N_b)/N,
\label{R}
\end{equation}
which appear as coefficients 
\begin{eqnarray}
A &=&(1+\Delta )(3+\Delta )-R(1-\Delta )(3-\Delta ), \\
B &=&(1+\Delta )(4+\Delta )-2-R(1-\Delta )(4-\Delta ), \\
C &=&2\Delta -R(1-\Delta ),
\end{eqnarray}
in a quadratic equation for $f^{*}$: 
\begin{eqnarray}
A\left( f^{*}\right) ^2-Bf^{*}+C=0. 
\end{eqnarray}
A similar equation is easily obtained for $h^{*}$, by replacing $(R,\Delta)\rightarrow (R^{-1},-\Delta )$. The solutions are readily found, though the expressions do not convey clear messages. To gain some insight, we consider two simple limits: $R=1$ and $\Delta =0$.\\

- Of course, the physics of $R=1$ is trivial, since the system consists of two identical populations ($\kappa _a=\kappa _b=\kappa $). Thus, $\langle k_{{\bf g}}\rangle ^{*}=\kappa $ and $f^{*}=h^{*}=1/2$. Here, $\Delta$ represents an arbitrary partition (perhaps based on other attributes) of a homogeneous population. We find easily $\langle k_{ab}\rangle ^{*}$ from considering the fraction of the average degree $\langle k_a\rangle ^{*}$, say, that are connected to the ``introverts,'' i.e., $\langle k_a\rangle^{*}N_b/N=\langle k_a\rangle ^{*}(1-\Delta )/2$. But, from equation (\ref{kakb}) this fraction is $\langle k_{ab}\rangle ^{*}/(1+\Delta )$, so that 
\begin{eqnarray}
\langle k_{ab}\rangle ^{*}=\kappa (1-\Delta ^2)/2. 
\end{eqnarray}
Naturally, this is also the result if we consider the fraction of $\langle k_b\rangle ^{*}$ connected to the extroverts. From this, it follows immediately $\langle k_{aa}\rangle ^{*}=\langle k_{bb}\rangle ^{*}=\kappa (1+\Delta ^2)/2$.
\\

- On the other hand, if the two populations are of the same size ($\Delta =0$) but distinct ($\kappa _a\neq \kappa _b$), the results are slightly less
trivial: 
\begin{eqnarray}
\langle k_{aa}\rangle ^{*} &=&\kappa _a-\langle k_{ab}\rangle ^{*},\quad
\langle k_{bb}\rangle ^{*}=\kappa _b-\langle k_{ab}\rangle ^{*}, \\
\langle k_{ab}\rangle ^{*} &=&\frac 12\left[ \kappa _a+\kappa _b-\sqrt{%
\kappa _a^2-\kappa _a\kappa _b+\kappa _b^2}\right] \,\,.
\end{eqnarray}
An interesting, and somewhat unexpected result is obtained for the limit $\kappa _b\ll \kappa _a$ ($R\ll 1$), in which the average number of links within the group of introverts, $\langle k_{bb}\rangle ^{*}$, approaches a finite fraction of $\kappa _b$. From the above, we have 
\begin{eqnarray}
\langle k_{ab}\rangle ^{*}=\frac 34\kappa _b\left( 1+O\left( R\right)\right), 
\end{eqnarray}
so that only a quarter of an introvert's connections are with other introverts. Again, the ratio $\moyenne{k_{ab}}^*/\moyenne{k_{bb}}^*$ appears to be an integer: $\theta_0=3$. The dynamics in this scenario may be referred to as ``frustrated'' and described as follows. In their attempt to establish large number of contacts, the extroverts are just as likely to hit upon the introverts and so, to increase $k_{ab}$. But this activity is not welcome by the introverts and links would be cut in response. If an introvert were connected to just as many cross-links as intra-group links, then both $k_{ab}$ and $k_{bb}$ are equally likely to decrease. But such a situation cannot be sustained, since only the cross-links are continually being formed (by the extroverts). Thus, there must be more cross-links for a sustainable balance to emerge. Examining the last panels of figure ($\ref{Graph1}$) again, we see that this ratio is quite consistent with the large $t$ behavior of simulation data. It would be useful if a simple and intuitive argument can be advanced to arrive at the ratio $\theta_0=\langle k_{ab}\rangle ^{*}/\langle k_{bb}\rangle ^{*}=3$.\\

With a good grasp of the final stationary state we obtain a general picture of the relaxation dynamics through the standard linear stability analysis. Defining the vector $\vec{\phi}(t)=(\delta_{aa},\delta_{ab},\delta_{bb})$, with $\delta_{\ggg\ggg'}\equiv\langle k_{\ggg\ggg'}\rangle -\langle k_{\ggg\ggg'}\rangle ^{*}$, one easily obtain the expression of the matrix ${\Bbb M}$ that governs its evolution: $\partial_t\vec{\phi}(t)=-{\Bbb M}\vec{\phi}(t)$. Three relaxation times emerge, given by the inverse of the eigenvalues of ${\Bbb M}$.
In the limit $\sigma>>1$, a careful examination reveals that the largest relaxation time $\tau$, which governs the asymptotic behavior, is of order one while the two remaining characteristic times scales as $1/\sigma$. The details of this analysis can be found in appendix. With more effort, we can compute the precise value of $\tau$ and compare the $e^{-t/\tau}$ tail to those in simulations. However, we doubt that such an exercise will provide any further insight, since we expect the agreement between a mean field approach and the fully stochastic problem to be good only at a qualitative level.

\section{Limitations of mean field theory}
From the previous section, we see that the simple minded, mean-field approach (with large $\sigma ,N$) provides a good qualitative picture of the evolution of our network. On the other hand, it is also clear that there are limitations, which are not easily anticipated. This section is devoted to a brief description of some of the shortcomings and speculations on the missing crucial ingredients.
\subsection{Drifts in the intermediate regime}
\begin{figure}[h] 
\begin{center}
\epsfxsize=10cm
\mbox{\epsfbox{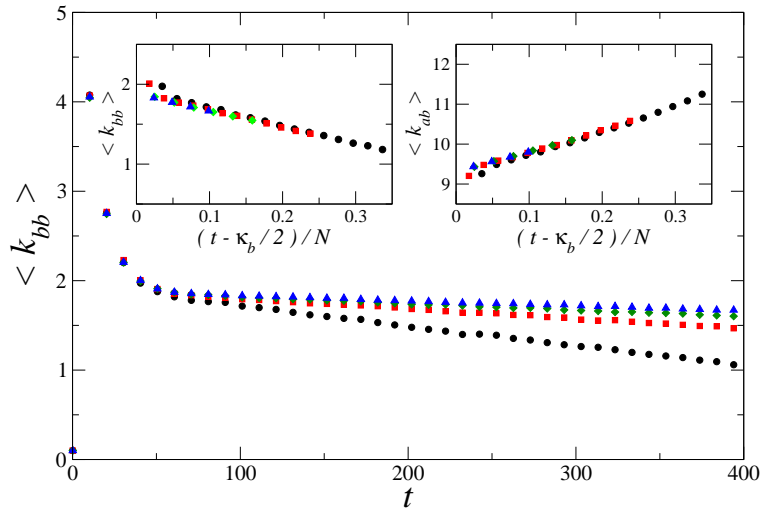}}
\caption{Average degree $\langle k_{bb}\rangle $ as a function of $t$ for different system sizes $N=1000$ (circles), $2000$ (squares), $3000$ (diamonds) and $4000$ (triangles). The data are extracted from simulations realized on $500$ realizations, for $\Delta =0$, $\kappa _a=500$, $\kappa _b=10$ and $\sigma =100$. Inset shows data collapse of $\langle k_{bb}\rangle$ and $\langle k_{ab}\rangle$ when plotted against $\left( t-t_1\right) /N$.}
\label{ZbbSize}
\end{center}
\end{figure}

\begin{figure}[h] 
\begin{center}
\epsfxsize=10cm
\mbox{\epsfbox{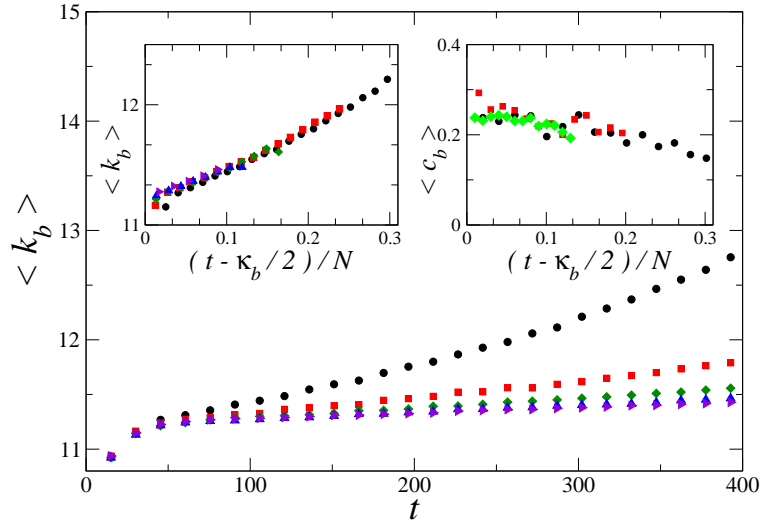}}
\caption{Average degree $\langle k_{b}\rangle $ as a function of $t$ for different system sizes $N=1000$ (circles), $2000$ (squares), $3000$ (diamonds), $4000$ (up triangles) and $5000$ (right triangles). The data are extracted from simulations realized on $500$ realizations, for $\Delta =0$, $\kappa _a=500$, $\kappa _b=10$ and $\sigma =100$. Inset shows data collapse of $\langle k_{b}\rangle$ and $\langle c_{b}\rangle$ when plotted against $\left( t-t_1\right) /N$.}
\label{ZbbSize2}
\end{center}
\end{figure}

From our simulation data, with $\sigma \gg 1$ and $N=1000,...,5000$, we observe a slow drift -- that scales with $1/N$ -- in all quantities associated with the introvert population, e.g., $\langle k_{ab}\rangle ,\langle k_{bb}\rangle $, etc. Most significantly, both $\langle k_b\rangle $ and $\langle c_b\rangle $ drift with $t$. But, in a mean field approximation, the latter probability (which is $\left\langle \left[ 1+\left( k_b/\kappa _b\right) ^\sigma \right]^{-1}\right\rangle $ ) is replaced by $h=\left[ 1+\left( \left\langle k_b\right\rangle /\kappa _b\right) ^\sigma \right] ^{-1}$ and so, becomes intimately related to $\left\langle k_b\right\rangle $. In particular, if $\sigma \gg 1$, small changes in $\left\langle k_b\right\rangle $ would induce drastically magnified (or suppressed) changes in $h$. Examining figures \ref{ZbbSize} and \ref{ZbbSize2}, it is clear that the drifts in these quantities are comparable, rather than $\partial _t\langle c_b\rangle \thicksim \sigma \partial _t\left\langle k_b\right\rangle $ (or $\partial _t\langle c_b\rangle \cong 0$). Such results are clear signals that correlations like $\langle c_b\left( k_b\right) \rangle -c_b\left( \langle k_b\rangle \right) $ are non-trivial. Even if we ignore such difficulties and examine the $O\left( 1/N\right) $ corrections in the mean field equations (\ref{EQA1}-\ref{EQA2}), we can show
that they cannot produce a consistent set of results that includes a linear $t$ dependence with $O\left( 1/N\right) $ coefficients. A more interesting behavior is the scaling of the restricted degree $\moyenne{k_{bb}}$ with the size $N$ of the population. By choosing the origin to be $t_1$ and using the scaled variable $\left( t-t_1\right) /N$, the data for all the $N$'s can be collapsed (shown in the inset in figures \ref{ZbbSize} and \ref{ZbbSize2}). Given that mean field theory cannot even predict a linear drift, we believe that an explanation of scaling can only emerge with a better understanding of the stochastics. To initiate further studies along these lines, one of us (RZ) has investigated the distribution of degrees for the simplest version of this model possible -- single population, $c\left( k\right) =\Theta \left( \kappa -k\right) $ -- and found it to be an exponential \cite{LSZ}. We believe that such non-trivial distributions will provide some the crucial ingredients needed to establish a theory for both the existence of these slow drifts and their scaling functions.
\subsection{Finite $\sigma $ effects and universality for large $\sigma$}
With moderate $\sigma $'s, the creation/destruction probabilities are rounded, so that mean field theory predicts smoother transitions between the regimes. This rounding is especially pronounced for the second transition, $t\thicksim t_2$ . As can be seen in figure ($\ref{Graph1}$), this transition is obscured even for $\sigma =10$, displaying only smooth, non-monotonic behavior. In other words, finite $\sigma $ rounding in the theory becomes significant when $\sigma $ drops below $50$. By contrast, the noise in the simulations completely masks this transition, even for $\sigma =100$. At the same time, (apart from a constant off-set) the three sets of data with $\sigma \gtrsim 5$ are essentially identical through this transition. Again, we view such differences between predictions and simulations as limitations of the mean field theory.\\
Another interesting aspect of finite $\sigma $ is `frustration.' Since everyone is more `flexible,' the extroverts/introverts can tolerate a finite level of cross-links being destroyed/created by the other group. As a result,  $\langle r_a\rangle ^{*}\equiv \langle k_a\rangle ^{*}/\kappa _a<1$ and $\langle r_b\rangle ^{*}\equiv \langle k_b\rangle ^{*}/\kappa _b>1$, as seen in simulations. Naturally, $\Delta \equiv (N_a-N_b)/N$ also plays a role. In general, the theory above predicts $f^{*},h^{*}\neq 1/2$. Given $\langle r_a\rangle ^{*}=\left( 1/f^{*}-1\right) ^{1/\sigma }$ etc., we see that the ratios will differ from unity for finite $\sigma $. We can gain some insight by rewriting equation ($\ref{EqualityCaCb}$) in a manifestly symmetric form: 
\begin{eqnarray}
\left[ \left( \langle r_a\rangle ^{*}\right) ^\sigma -\Delta \right] \left[
\left( \langle r_b\rangle ^{*}\right) ^\sigma +\Delta \right] =1-\Delta ^2.
\end{eqnarray}
Clearly, the two ratios lie on opposite sides of unity. By computing their precise values, we see that they agree reasonably well with data (e.g., for $t=300$ in figure $\ref{Graph1}$). In this connection, we note that the agreement is much better for intra-group links (e.g., $\langle k_{bb}\rangle^{*}$) than for the cross-links. Presumably, this aspect is also a limitation of mean field theory.
\section{Summary and outlook}
Motivated by the competing dynamics of different social groups, we analyzed the birth, growth and relaxation of a network formed by two populations. The main distinction between these two is their preference for having different degrees: $\kappa _a$ and $\kappa _b$. With a society of extroverts and introverts in mind, we considered mainly systems with $\kappa _a=100$ and $\kappa _b=10$. In our study, we focus only on the evolution of the network,
i.e., the connections between individuals. In particular, during each time-step of our simulation (defined by the random selection of $N$ individuals), we count the degree ($k$) of each node to determine the probability to create or destroy a link ($c\left(k\right) $, $1-c\left( k\right) $). In this work, we chose $c\left( k\right)=1/[ 1+\left( k/\kappa \right)^\sigma ] $, which is effectively a Fermi-Dirac function, starting at $c\left( 0\right) =1$, dropping monotonically with $k$ through $1/2$ over a range controlled by $\sigma$. In this manner, if the degree of the selected node is too low, it creates a link with a randomly picked new ``friend.'' On the other hand, if its degree is too high, a random chosen link is cut.

Performing computer simulations, we evolve an initially unconnected set of $N=N_a+N_b$ individuals, with a variety of ${\sigma}$'s and $N$'s. All data collected reveal three time regimes, characterized by non-monotonic
behavior of the introverts' degrees. In the early regime, all individuals are creating links, so that the connectivity in both groups grows linearly. The onset of the intermediate regime is signaled by the introverts reaching their preferred degree while the extroverts remain unsatisfied. Not surprisingly, the cross-links continue to be created by the latter. The introverts struggle to cut links, but, doing so at random (in our model), find themselves with more cross-links than a sustainable value. In other words, the cross-links overshoot, while the introvert-introvert links undershoot. Finally, the late regime is characterized by the extroverts having reached their preferred degree, so that the final, stationary state is reached through the redistribution between intra-group links and cross-links. To describe these scenarios with more than words, we turn to various theoretical approaches.

A full description of the system can be formulated by a master equation for $P\left( \left\{ \mu _{i,j}\right\} ,t\right) $, where $\mu _{i,j}=0$/$\mu _{i,j}=1$ corresponds to the absence/presence of a link between individuals $i$ and $j$. Though we can write such an equation, we cannot solve it. Instead, we consider an approximate formulation, using only three ``global''
variables: $\alpha $ ($\beta $), the total number of links between the extroverts (introverts) and $\chi $, the total number of cross-links. Even in this reduced phase space, we cannot solve the master equation exactly. To make progress, we exploit a standard mean field approach, writing equations of evolution for the {\em average} degrees within each population ($\langle k_{aa}\rangle $ and $\langle k_{bb}\rangle $) and those associated with cross-links: $\langle k_{ab}\rangle $. The emergence of the three time regimes can now be predicted and qualitatively good agreement between theory and simulations are found. Surprisingly, in the case $N_a=N_b$, the ratio $\theta_0$ of the restricted degrees $\moyenne{k_{ab}}$ and $\moyenne{k_{bb}}$ is given by an integer in the asymptotic regime of both transient and stationary states. Of course, there are limitations to this crude theory, some of which we duly noted.

Our study should be regarded as the first step towards a fully interacting adaptive network, in which the individuals are characterized by some states (e.g., opinions, susceptible/infected, etc.) as well. In this work, only the links are endowed with dynamics. Even so, there are many unanswered questions which deserves being pursued. Exploring ways to go beyond mean field theory, by studying the full distribution in $k$, is just an example. Work on this aspect is in progress and will be presented elsewhere \cite{LSZ}. Having more than two groups is another. In reality, every individual will have its own preferred degree ($\kappa _i$) or flexibility ($\sigma _i$), which generates an interesting problem with quenched disorder. Needless to say, we can study further variations, each incorporating more aspects of real social networks.

If we endow individuals with states and their dynamics, then the vistas are even wider. Since one of our motivations for the present study is how epidemics can be affected by social behavior, let us provide a simple example by considering the epidemiological $SIS$ model on such a dynamic network. See \cite{Keeling-Eames-2005} for a review of the basis of epidemiological theory and network theory. In this framework, models that involve adaptive networks have been considered in \cite{Gross-2006,Leah-2008,Leah-2010}, where the networks
``rewires'' by having nodes shifting their connections from one friend to another. If we admit the simplest psychology and let the preferred degrees $\kappa$ depend appropriately on the level of the epidemic (through the total fraction of infected individuals, say), we can expect oscillations to occur. Promising preliminary results have been obtained \cite{TP} for a much simpler version. In that model, mean field theory predicts only fixed points with no limit cycles (though underdamped oscillations can occur). Simulations which include stochastics show only a simple stationary distribution, although a signature of the feedback induced oscillations can be found in the autocorrelations (and its power spectrum \cite{Alonso-2007,Simoes-2008}). In this regard, we believe that novel and interesting phenomena will emerge if feedback is introduced into the system by coupling of the dynamics of the nodes with that of the links. If realized, such a system will have its own natural frequencies (periods) that depend only on the system parameters. How these ``internal'' periods interact with those from ``external'' drives (e.g., seasonal variations of infectious diseases) will surely be of interest in the modeling of epidemics in interacting populations.

\section{Appendix: Linear stability analysis}
For our analysis, we consider the $N\rightarrow \infty $ limit only, i.e., equations  (\ref{dQ/dt}), (\ref{MEANFIELDX}), (\ref{MEANFIELDX2}) and (\ref{MEANFIELDX3}). For convenience of notation, let us define 
\begin{eqnarray}
x,y,z \equiv \langle k_{aa}\rangle ,\langle k_{ab}\rangle ,\langle k_{bb}\rangle \quad\textrm{and}\quad \mu _{\pm } \equiv 1/\left( 1\pm \Delta \right),
\end{eqnarray}
so that 
\begin{eqnarray}
r_a &=&\frac{x+\mu _{+}y}{\kappa _a}, \quad \quad r_b=\frac{z+\mu _{-}y}{\kappa _b}.
\end{eqnarray}
Therefore, for {\em arbitrary} $f\left(r_a\right) $ and $h\left(r_b\right)$, we have 
\begin{eqnarray}
\partial_xf &=&f'/{\kappa _a},\quad \quad \partial_yf=\mu _{+}\partial_xf, \\
\partial_zh &=&{h'}/{\kappa _b},\quad \quad \partial_yh=\mu _{-}\partial_xh.
\end{eqnarray}
A useful guide is to keep in mind our {\em specific} case, in which $-f^{\prime }=\sigma f\left( 1-f\right)$ and $-h^{\prime }=\sigma h\left( 1-h\right)$ showing explicitly the $O\left( \sigma \right) $ nature of these derivatives. Let us write the equations for $\partial _tx$ and $\partial _tz$ as 
\begin{eqnarray}
\partial _tx=Xf-X_1\quad \textrm{and}\quad \partial _tz=Zh-Z_1,
\end{eqnarray}
where $X \equiv 1+\Delta +X_1$ and $Z \equiv 1-\Delta +Z_1$ with
\begin{eqnarray}
X_1\equiv \frac{2x}{x+\mu _{+}y}, \quad\textrm{and} \quad Z_1\equiv \frac{2z}{z+\mu _{-}y}.
\end{eqnarray}
For $\partial _ty$, one obtains
\begin{eqnarray}
\partial _ty=\frac{1+\Delta }2\left( 4-X\right) f+\frac{1-\Delta }2\left(
4-Z\right) h+\frac{1+\Delta }2X_1+\frac{1-\Delta }2Z_1-2. \nonumber\\
\end{eqnarray}
Defining the vector
\begin{equation}
\vec{\phi}(t)\equiv \left( 
\begin{array}{c}
\langle k_{aa}\rangle -\langle k_{aa}\rangle ^{*} \\ 
\langle k_{ab}\rangle -\langle k_{ab}\rangle ^{*} \\ 
\langle k_{bb}\rangle -\langle k_{bb}\rangle ^{*}
\end{array}
\right),
\label{phi}
\end{equation}
and linearizing equations (\ref{MEANFIELDX}), (\ref{MEANFIELDX2}) and (\ref{MEANFIELDX3}) around the fixed point, we write 
\begin{eqnarray}
\partial _t\vec{\phi}(t)=-{\Bbb M}\vec{\phi}(t)\,\,.  \label{M}
\end{eqnarray}
We see that every element of ${\Bbb M}$ will contain $f^{\prime }$ or $h^{\prime }$ (or both), which are $O\left( \sigma \right)$. Denoting this order by ${\Bbb M}_0$ we have 
\begin{eqnarray}
{\Bbb M}_0=\left( 
\begin{array}{ccc}
u & \mu _{+}u & 0 \\ 
\tilde{u} & \mu _{+}\tilde{u}+\mu _{-}\tilde{v} & \tilde{v} \\ 
0 & u _{-}v & v
\end{array}
\right), 
\end{eqnarray}
where 
\begin{eqnarray*}
u &\equiv &X^{*}\left( -f^{\prime }\right) ^{*}/\kappa _a,\quad \tilde{u}\equiv \frac{1+\Delta }2\left( 4-X^{*}\right) \left( -f^{\prime
}\right) ^{*}/\kappa _a, \\
v &\equiv &Z^{*}\left( -h^{\prime }\right) ^{*}/\kappa _b,\quad \tilde{v}\equiv \frac{1-\Delta }2\left( 4-Z^{*}\right) \left( -h^{\prime
}\right) ^{*}/\kappa _b,
\end{eqnarray*}
denote the above functions at the fixed point.  In the large $\sigma $ limit of interest here, the simple expectation is that all eigenvalues will be proportional to $\sigma $. However, this would lead to an  {\em instantaneous} relaxation into the steady state (i.e., with $O\left( 1/\sigma \right)\rightarrow 0$ decay time). A more careful examination reveals that ${\Bbb M}_0$ has a zero eigenvalue, with eigenvectors 
\begin{eqnarray}
\left\langle 0\right| \equiv \left( 
\begin{array}{ccc}
\frac 1{2\mu _{+}}\left( 1-\frac 4{X^{*}}\right), & 1, & \frac 1{2\mu _{-}}%
\left( 1-\frac 4{Z^{*}}\right)
\end{array}
\right) \quad \textrm{and}\quad \left| 0\right\rangle \equiv \left( 
\begin{array}{c}
-\mu _{+} \\ 
1 \\ 
-\mu _{-}
\end{array}
\right),
\end{eqnarray}
with 
\begin{eqnarray}
\left\langle 0\right| \left. 0\right\rangle =\frac 2{X^{*}}+\frac 2{Z^{*}}.
\end{eqnarray}
Since both its trace and its co-factor (respectively, $v +u +\mu _{+}\tilde{u}+\mu _{-}\tilde{v}$ and $u v +\mu _{+}\tilde{u}v+\mu _{-}\tilde{v}u $) are non-zero, the other two eigenvalues are $O\left( \sigma \right) $. Expecting both to be positive on physically grounds, we do not pursue their precise values.\\
Given that we have the eigenvectors at $O\left( \sigma \right) $, we can compute the eigenvalue at $O\left( 1\right) $ by standard perturbation theory. Calling that part of the matrix ${\Bbb M}_1$, the $O\left( 1\right)$ eigenvalue is given by standard perturbation theory: 
\begin{eqnarray}
\lambda =\left\langle 0\right| {\Bbb M}_1\left| 0\right\rangle \frac{X^{*}Z^{*}}{2\left( X^{*}+Z^{*}\right) }, 
\end{eqnarray}
where ${\Bbb M}_1$ is the $O\left( 1\right) $ terms in ${\Bbb M}$. The decay of any quantity that has overlap with $\left| 0\right\rangle $ will be {\em independent }of $\sigma $ in the large $\sigma $ limit.
\section*{Acknowledgements}
We thank S. Dorosz, B. Schmittmann, and L.B. Shaw for illuminating discussion. This research is supported in part by the US National Science Foundation through DMR-0705152.

\end{document}